\documentstyle[epsfig,aps,prl,multicol]{revtex}
\begin{document}

\title{Variational procedure and generalized Lanczos recursion for
small-amplitude classical oscillations}

\author{E.V. Tsiper}

\address{Department of Physics, SUNY at Stony Brook, Stony Brook, NY
11794}

\date{To appear in JETP Letters {\bf 70}, 11, 740 (1999)}

\maketitle

\begin{multicols}{2}
 [ \begin{abstract}
 Variational procedure is developed that yields lowest frequencies of
small-amplitude oscillations of classical Hamiltonian systems.
Genuine Lanczos recursion is generalized to treat related
non-Hermitian eigenvalue problems.
 \end{abstract} ]

Normal modes $\xi$ and frequencies $\omega$ of small oscillations of a
classical system near the equilibrium are determined by the secular
equation \cite{landau1}

\begin{equation}
\omega^2M\xi=K\xi,
\label{mech}
\end{equation}
 where $M$ and $K$ are $N\times N$ symmetric positive definite
matrices of mass coefficients and spring constants respectively.  In
many applications the number $N$ of degrees of freedom is large, while
only a few lowest frequencies are of interest \cite{parlett}.
Equation (1) represents a problem more complex than a regular
symmetric eigenvalue problem, unless $M$ or $K$ is diagonal.

Equation (\ref{mech}) can be transformed into the Hamiltonian form by
introducing canonical momentum $\eta=\omega M\xi$:

\begin{equation}
K\xi=\omega\eta,\ \ \ \ \ \  T\eta=\omega\xi,
\label{mech2}
\end{equation}
 where $T=M^{-1}$.  Thus, the frequencies of the normal modes are the
eigenvalues of a $2N\times2N$ matrix

\begin{equation}
\left(\begin{array}{cc}
0 & T \\
K & 0 
\end{array}\right)
\label{matr}
\end{equation}

The spectrum of this matrix consists of pairs $\pm\omega$, since
$(\xi,-\eta)$ is also a solution of (\ref{mech2}) that corresponds to
$-\omega$.  The lowest frequency $\omega_{\min}$ is the lowest {\em
positive} eigenvalue of the matrix (\ref{matr}).

Although the eigenvalues of the matrix (\ref{matr}) are always real,
the matrix itself is non-Hermitian, unless $K=T$.  Therefore, its
diagonalization poses a formidable task.  The major problem is that no
general minimum principle exists that yields eigenvalues of arbitrary
diagonalizable non-Hermitian matrices.  This does not allow to
formulate a variational procedure similar to the Rayleigh-Ritz
procedure for Hermitian matrices.  If $K=T$, the matrix (\ref{matr})
is Hermitian, and its positive eigenvalues coincide with those of $K$
and $T$.

As known from quantum mechanics, the lowest eigenvalue
$\epsilon_{\min}$ of a Hermitian matrix $H$ can be obtained from the
minimum principle

\begin{equation}
\epsilon_{\min}=\min_{\{\psi\}}\frac{(\psi H\psi)}{(\psi\psi)}.
\label{rayleigh}
\end{equation}

The minimum is to be searched over all vectors $\psi$.  The Ritz
variational procedure is an approximation when the set $\{\psi\}$ in
(\ref{rayleigh}) is restricted to some subspace ${\cal K}$ of
dimension $n<N$.  The best approximation to $\epsilon_{\min}$ in the
sense of (\ref{rayleigh}) is obtained as the lowest eigenvalue of the
$n\times n$ Rayleigh matrix $\widetilde H$, obtained by projection of
$H$ onto ${\cal K}$.

The special paired structure of the matrix (\ref{matr}) makes it
possible to generalize (\ref{rayleigh}) such as to yield
$\omega_{\min}$.  In fact,

\begin{equation}
\omega_{\min}=\min_{\{\xi,\eta\}}
  \frac{(\xi K\xi)+(\eta T\eta)}{2\left|(\xi\eta)\right|}.
\label{rr}
\end{equation}

The minimum is to be searched over all possible phase space
configurations $\{\xi,\eta\}$.  Before providing the proof to this
equation let me point out some of its features.

First, it states that $\omega_{\min}$ is the minimum harmonic part of
the total energy, $(\xi K\xi)/2+(\eta T\eta)/2$, over the phase space
configurations normalized by $(\xi\eta)=1$.  Since $K$ and $T$ are
both positive definite, the right-hand side is strictly positive and
so is $\omega_{\min}$.  Second, equation (\ref{rr}) is symmetric in
$K$ and $T$, according to the nature of the problem.  When $K=T$ the
minimum is achieved at $\xi=\eta$, and (\ref{rr}) becomes the same as
(\ref{rayleigh}).

Note that the functional in (\ref{rr}) has no maximum, since the
denominator can be made arbitrarily small.  The global minimum,
however, always exists.  This is not obvious, since a set of all pairs
of vectors with $(\xi\eta)=1$ is not compact.  Indeed, say, any vector
orthogonal to $\eta$ can be added to $\xi$, making $|\xi|$ arbitrarily
large.  However, the functional in (\ref{rr}) grows indefinitely in
this case, such that the global minimum is achieved at finite $|\xi|$
and $|\eta|$.

Variation of (\ref{rr}) with respect to $\xi$ and $\eta$ yields
equations (\ref{mech2}).  Thus, the solutions of (\ref{mech2}) are the
stationary points of (\ref{rr}).  The global minimum (\ref{rr}),
therefore, gives indeed $\omega_{\min}$.  The singularity in the
denominator poses no problem, since it corresponds to infinitely large
values of functional, while near the minimum it is analytic.

Minimum principle (\ref{rr}) can, in fact, be obtained from the
Thouless minimum principle \cite{thouless61}, derived for
non-Hermitian matrices that appear in random phase approximation
(RPA).  Equation (\ref{rr}) transforms into the Thouless minimum
principle by substitution: $A=(K+T)/2$, $B=(K-T)/2$, $x=(\xi+\eta)/2$,
and $y=(\xi-\eta)/2$.

Variational procedure similar to the Rayleigh-Ritz procedure can be
formulated if coordinates $\xi$ and momenta $\eta$ in (\ref{rr}) are
restricted to some subspaces ${\cal U}$ and ${\cal V}$ of dimension
$n$, respectively.

Let $\{\xi_i\}$ and $\{\eta_i\}$ be two sets of vectors that span
${\cal U}$ and ${\cal V}$, such that $(\xi_i\eta_j)=\delta_{ij}$.
Expanding $\xi=\sum u_i\xi_i$, $\eta=\sum v_i\eta_i$ and varying
(\ref{rr}) with respect to $u_i$ and $v_i$, we find the latter to obey
a $2n\times2n$ eigenvalue equation

\begin{equation}
\left(\begin{array}{cc}
0 & \widetilde T \\
\widetilde K & 0 
\end{array}\right)
\left(\begin{array}{c}
u \\
v 
\end{array}\right)
  = \widetilde\omega
\left(\begin{array}{c}
u \\
v 
\end{array}\right),
\label{rrritz}
\end{equation}
 with $\widetilde K_{ij}=(\xi_i K\xi_j)$ and $\widetilde
T_{ij}=(\eta_i T\eta_j)$.  Equation (\ref{rrritz}) generalizes
Hermitian Rayleigh-Ritz eigenvalue equation for $\widetilde H$.  It
has $2n$ solutions $\pm\widetilde\omega$, the lowest positive of which
gives the best approximation to $\omega_{\min}$ in the sense of
equation (\ref{rr}).

Krylov subspace \cite{parlett} for the matrix (\ref{matr}) can be
constructed by acting with it many times on an arbitrary vector
$(\xi_1,\eta_1)$:

\begin{equation}
\left(\begin{array}{c}
\xi_1 \\
\eta_1 
\end{array}\right),\ 
\left(\begin{array}{c}
T\eta_1 \\
K\xi_1 
\end{array}\right),\ 
\left(\begin{array}{c}
TK\xi_1 \\
KT\eta_1 
\end{array}\right),\ ...
\label{seq}
\end{equation}

The subspace that spans first $n$ vectors of this sequence has the
property of approximating an invariant subspace of (\ref{matr}).
Thus, it is natural to expand approximation to an eigenvector of
(\ref{matr}) as a linear combination of these vectors.  In other
words, the natural choice for the subspaces ${\cal U}$ and ${\cal V}$
for the variational procedure described above are the subspaces ${\cal
U}_n$ and ${\cal V}_n$ that span the upper and lower components of
first $n$ vectors of (\ref{seq}).

In order to implement the variational procedure, it is necessary to
construct a biorthogonal basis $\{\xi_i,\eta_i\}$, $i=1,...,n$ in
${\cal U}_n$ and ${\cal V}_n$ and compute matrix elements of
$\widetilde K$ and $\widetilde T$.  Both tasks can be performed
simultaneously using the following recursion:

\begin{mathletters}
\begin{eqnarray}
\xi_{i+1}&=&\beta_{i+1}^{-1}(T\eta_i-\alpha_i\xi_i-\beta_i\xi_{i-1})
\label{rrlanc1}\\
\eta_{i+1}&=&\delta_{i+1}^{-1}(K\xi_i-\gamma_i\eta_i-\delta_i\eta_{i-1}).
\label{rrlanc2}
\end{eqnarray}
\label{rrlanc}
\end{mathletters}
 The four coefficients $\alpha_i$, $\beta_i$, $\gamma_i$, and
$\delta_i$ are to be chosen at each step $i$ such as to make
$\xi_{i+1}$ orthogonal to $\eta_i$ and $\eta_{i-1}$, and $\eta_{i+1}$
--- orthogonal to $\xi_i$ and $\xi_{i-1}$.  This appears to be enough
to ensure global biorthogonality $(\xi_i\eta_j)=\delta_{ij}$.

Indeed, assume biorthogonality to hold up to step $i$.  Multiplying
(\ref{rrlanc1}) by $\eta_j$, $j<i-1$, we have
$(\eta_j\xi_{i+1})\propto(\eta_jT\eta_i)=(\eta_iT\eta_j)=0$ due to
Hermiticity of $T$ and the fact that $T\eta_j$ is a linear combination
of all $\xi_k$ with $k\leq j+1<i$.  Thus, the biorthogonality also
holds for the step $i+1$.

Multiplying (\ref{rrlanc1}) by $\eta_{i-1}$, $\eta_i$, and
$\eta_{i+1}$ and using biorthogonality, we get $(\xi_i\eta_i)=1$,
$\widetilde K_{ii}=\alpha_i$, and $\widetilde K_{i,i-1}=\widetilde
K_{i-1,i}=\beta_i$.  Similarly, $\widetilde T_{ii}=\gamma_i$ and
$\widetilde T_{i,i-1}=\widetilde T_{i-1,i}=\delta_i$.  All other
matrix elements of $\widetilde K$ and $\widetilde T$ vanish.

The recursion (\ref{rrlanc}) is a straightforward generalization of
the Hermitian Lanczos recursion \cite{parlett,lanczos}

\begin{equation}
\psi_{i+1}=\beta_{i+1}^{-1}(H\psi_i-\alpha_i\psi_i-\beta_i\psi_{i-1})
\label{lanc}
\end{equation}
 applicable to any Hermitian matrix $H$.  When $K=T$ and
$\xi_1=\eta_1$ both equations (\ref{rrlanc}) coincide with each other
and with equation (\ref{lanc}), up to the notation.

As in the case of the Hermitian Lanczos algorithm, several lowest
frequencies can be found one by one by projecting the $\xi-$ and
$\eta$-components of converged eigenvectors out of ${\cal V}_n$ and
${\cal U}_n$ subspaces respectively.

The method was tested on a set of large sparse random matrices of the
form (\ref{matr}).  Symmetric matrices $T$ and $K$ were generated to
have an average of 40 randomly distributed and randomly positioned
matrix elements in each row.  Both $K$ and $T$ were shifted by an
appropriate constant to ensure positive definiteness.  Figure 1
demonstrates the convergence results for a matrix of the size
$2N=200000$.

\begin{center}
\begin{figure}
\epsfig{file=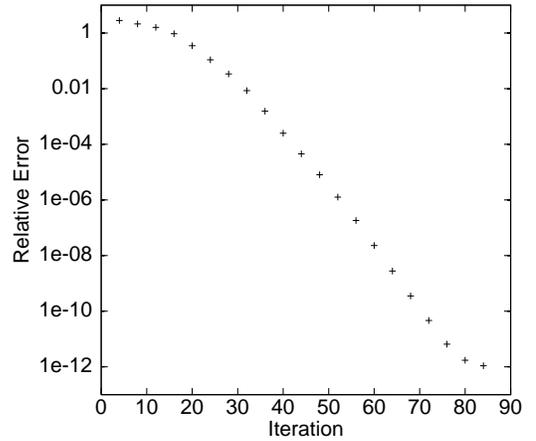,width=7cm}
 \protect\caption{Convergence of the generalized Lanczos algorithm for
a random matrix of the form (\ref{matr}) and size $2N=200000$}
 \end{figure}
\end{center}

For smaller matrices up to $2N=2000$, where it was possible to obtain
all eigenvalues with regular methods, the present method has converged
to the true lowest frequency in all instances.

In conclusion, the method is proposed that generalizes Rayleigh-Ritz
variational procedure and Lanczos recursion to the case of
non-Hermitian matrices of the form (\ref{matr}), that determine normal
modes and frequencies of small-amplitude oscillations of Hamiltonian
systems.

Equations (\ref{mech2}) have numerous applications beyond purely
mechanical problems.  Schroedinger equation in non-orthogonal basis
represents a generalized symmetric eigenvalue problem similar to
(\ref{mech}).  RPA and other time-dependent techniques in nuclear
physics and quantum chemistry lead to the equations similar to
(\ref{mech2}) \cite{thouless61,blaizot}.  At last, eigenvectors of
so-called Hamiltonian matrices, to which (\ref{matr}) is a special
case, solve the nonlinear algebraic Riccati equation which appears in
the theory of stability and optimal control \cite{riccati}.

I would like to acknowledge numerous enlightening discussions with
Vladimir Chernyak during my appointment at the University of
Rochester.

\end{multicols}
\end{document}